\documentclass[a4paper,12pt]{article}
\usepackage[utf8]{inputenc}
\usepackage{cancel}
\usepackage{ulem}
\usepackage{amsfonts}
\usepackage{amssymb}
\usepackage{graphicx}
\usepackage{amsmath}
\usepackage{enumerate}
\usepackage{mathtools}
\usepackage{subfig}
\usepackage{color}
\usepackage{float}
\usepackage{here}
\usepackage{cite}
\usepackage{mathrsfs}
\usepackage{float,epsfig}
\usepackage{dcolumn}
\usepackage{graphicx}
\usepackage{bm}
\usepackage{amsmath,amssymb,amsthm}
\usepackage[colorlinks=true,linkcolor=blue,citecolor=red]{hyperref}
\usepackage{multirow}

\newcommand{\be}{\begin{equation}}
\newcommand{\ee}{\end{equation}}
\newcommand{\bea}{\setlength\arraycolsep{2pt} \begin{eqnarray}}
\newcommand{\eea}{\end{eqnarray}}

\def\0{{\sst{(0)}}}
\def\1{{\sst{(1)}}}
\def\2{{\sst{(2)}}}
\def\3{{\sst{(3)}}}
\def\4{{\sst{(4)}}}
\def\5{{\sst{(5)}}}
\def\6{{\sst{(6)}}}
\def\7{{\sst{(7)}}}
\def\8{{\sst{(8)}}}
\def\sst#1{{\scriptscriptstyle #1}}

\makeatletter \@addtoreset{equation}{section}

\begin{document}

\title{ \textbf{\Large Dyonic Objects and Tensor Network Representation}}
\author{ A. Belhaj$^{1}$, Y. El Maadi$^{1}$, S-E. Ennadifi$^{2}$, Y. Hassouni%
$^{1}$, M. B. Sedra$^{3,4}$\thanks{%
Authors in alphabetical order} \hspace*{-8pt} \\
{\small $^1$ D\'epartement de Physique, Equipe des Sciences de la mati\`ere
et du rayonnement, ESMaR, }\\
{\small Facult\'e des Sciences, Universit\'e Mohammed V de Rabat, Rabat,
Morocco}\\
{\small $^{2}$ LHEP-MS, D\'epartement de Physique, Facult\'{e} des Sciences}%
\\
{\small Universit\'{e} Mohammed V de Rabat, Morocco}\\
{\small $^{3}$D\'{e}partement de Physique, LabSIMO, Facult\'{e} des
Sciences, Universit\'{e} Ibn Tofail }\\
{\small K\'{e}nitra, Morocco} \\
{\small $^4$ Facult\'e des Sciences et Techniques d'Errachidia, Universit\'e
Moulay Ismail,}\\
}
\maketitle

\begin{abstract}
{\noindent}

Motivated by particle physics results, we investigate certain dyonic
solutions in arbitrary dimensions. Concretely, we study the stringy
constructions of such objects from concrete compactifications. Then, we
elaborate their tensor network realizations using multistate particle
formalism. \newline
\end{abstract}

\tableofcontents

%

\newpage 

\section{Introduction}

Since the development of string theory framework and its efficient technique
for studying non-perturbative phenomena, a certain number of relevant links
with other theories have been comprehended \cite{1,2}. Specially, in such a
framework, the formerly familiar group of solitonic p-brane solutions of
Type II supergravities in 10 dimensions are known to be characterized in
terms of D-branes \cite{2,3}. Non-perturbative properties of supersymmetrric
Yang-Mills theory, black holes entropy, dyons, magnetic monopoles and a
number of appealing issues in various dimensions have been handled in the
context of D-brane objects and their dynamics \cite{4,5}. All these relevant
connections have promoted such a string theory aspect to one of the most
important and promising side to be explored. It is known that in higher
dimensions many physical quantities are reinterpreted from the usual
four-dimensional spacetime viewpoint as various faces of possible one
quantity. In particular, it has been shown that the single electric and
magnetic charges in ten dimensions could be viewed from the 4-dimensional
non-compact spacetime standpoint as dyonic charges, being particles carrying
simultaneous existence of the electric and the magnetic charges \cite{6}. A
dyon with a zero electric charge is usually referred to as a magnetic
monopole \cite{7,8}. These objects are hypothetical particles predicted in
many extended models  and grand unified theories \cite{9,10}. Indeed, besides to
magnetic monopole solutions, dyonic solutions appear in extended
Yang-Mills-Higgs modeling  taking places in theories going beyond the ordinary
physics associated with material points. It is therefore interesting to see
how the D-brane framework automatically encodes this feature by using
stringy constructions and computations.

Although there appears to have been no precedent concrete experimental
search for dyonic objects, it has been now become possible given the
progress made in the direct search of magnetic monopoles at accelerators
which has a long history \cite{11}. Actually, seen that dyons possess an
electric charge that can, in principle, be significant, with the recent LHC
results. For instance ATLAS and MoEDAL results have reported limits on
monopole productions as well as on stable objects with electric charge \cite{12,13}. Thus, it turns out that LHC could have a right place for the search
of dyons.

The aim of this work is contribute to this field by investigating stringy
dyonic objects and the corresponding tensor network realizations. Inspired
and motivated by known results obtained from  non-trivial theories including
string theory, we first provide the origin and certain related concepts
associated with such objects. After that,  we present a general framework of
dyonic solutions in the brane structure by showing how they are naturally
built from inspired stringy compactifications on certain complex manifolds.
Then, we reveal that such dyonic solutions involve tensor network
representations.

The organization of this work is as follows. In section 2, we present
shortly the origin and the related concepts of the dyonic objects motivated
by known results. In section 3, we elaborate a general framework of dyonic
solutions in the brane structure and show how are naturally obtained in
various dimensions within stringy scheme compactifications. In section 4, we
investigate the dyonic solution constructions using the tensor network
formalism and reveal some dyonic object characteristics. The last section  is
devoted to some concluding remarks and open questions.

\section{Motivations of dyonic objects}

Before dealing with the  dyonic objects associated with the electromagnetic
duality in higher dimensions, we would like to shed light on some related
concepts. Indeed, a similar one goes back to the neutrino discovery, where
the proton and the neutron can be regarded as two states of a single
particle motivated by the observation that they have approximately equal
masses $m_{p}\simeq m_{n}=m$ \cite{14}, which in turn conducted, according
to the mass-energy equivalence, to an energy degeneracy of the underlying
interaction. This mass degeneracy led then to the existence of a
corresponding symmetry whose the interaction obeys. That is to say, these
two particles have an identical behavior under the underlying interaction
and that their charge content is their solely difference. In such a case,
where these two particles are to be viewed as two linearly independent
states of the same particle, i.e., nucleus $N$, it is unexceptional to
depict them in the form of a two component vector like
\begin{equation}
N:%
\begin{pmatrix}
\underline{p} \\[3mm]
\underline{n}%
\end{pmatrix}%
_{m_{\underline{p},\underline{n}}}
\end{equation}%
where $\underline{p}$ and $\underline{n}$ refer to the proton and the
neutron particles making such a nucleus state. More fundamentally, analogous
to the spin-up and  the spin-down states of a spin-one half particle, the concept
of isospin symmetry is introduced and is governed by an $SU(2)_{I}$ group
rotating doublet components, i.e., proton and neutron, into each other in
abstract isospin space. In particular, in the modern formulation, the
isospin is defined as a vector quantity in which up and down quarks have a
value of $1/2 $, with the 3rd-component being $+1/2$ for up quarks, and $%
-1/2 $ for down quarks. In this picture, denoting the total isospin $I$ and
its 3rd component $I_{3}=\pm 1/2$, an up-down quarks pair can be assembled
in a state of the total isospin $1/2$ as

\begin{equation}
\begin{pmatrix}
\underline{u} \\[3mm]
\underline{d}%
\end{pmatrix}%
_{I=1/2}
\end{equation}%
where, again, $\underline{u}$ and $\underline{d}$ refer to the up and down
quarks forming such a doublet state. Albeit such a particle state
classification is a good approximated one with a symmetry dealing with
nuclear interactions. It remains a useful and large concept. In fact, one
could go further and extend this view to the electromagnetic charge where
each particle P could be represented according to its charge content by the
doublet
\begin{equation}
\mbox{P}:%
\begin{pmatrix}
Q_{e} \\[3mm]
P_{m}%
\end{pmatrix}%
_{Q_{em}},
\end{equation}%
whose components are the  electric and magnetic charges of the involved particle. At first
sight, this picture will seem to be strange or even inconsistent with the
symmetry of the electromagnetic interaction as dictated by the corresponding
$U(1)_{Q_{em}}$ group. However, this apparent inconsistency could be simply
alleviated by assuming a hidden symmetry group associated with the magnetic
charge. Within the the standard model framework, a simple and economic way
to do so is the consider an extended electroweak symmetry $SU(2)_{L}\times
U(1)_{e}$ $\times $ $U(1)_{m}$ involving both electric $U(1)_{e}$ and
magnetic $U(1)_{m}$ symmetries being broken at a certain high energy
resulting in the standard low energy electromagnetic symmetry $U(1)_{Q_{em}}$%
. In this picture, and at a certain energy scale, the particle charge
representation in $ref$ gives rise to the following physical objects, for
instance
\begin{equation}
\begin{pmatrix}
0 \\
0%
\end{pmatrix}%
_{0},\quad
\begin{pmatrix}
Q_{e} \\
0%
\end{pmatrix}%
_{Q_{e}},\quad
\begin{pmatrix}
0 \\
P_{m}%
\end{pmatrix}%
_{Q_{m}},\quad
\begin{pmatrix}
Q_{e} \\[3mm]
P_{m}%
\end{pmatrix}%
_{Q_{em}}.
\end{equation}%
These four objects are nothing but all known possible particles, discovered
and hypothetical. In particular, the first \ neutral doublet is nothing but
the neutrino, being neutral. The second doublet corresponds to the ordinary
charged particles, the third doublet is that of theoretical magnetic
monopoles with a single magnetic charge, while the four doublet is associated with
to a dyonic particle carrying both electric and magnetic charges. For more
clarity, we list them in Tab.(\ref{tab0}).
\begin{table}[th]
\begin{center}
\begin{tabular}{|c|c|c|c|}
\hline
object & $Q_{e}$ & $P_{m}$ & nature \\ \hline
neutrinos & no & no & real \\ \hline
ordinary (electrically) charged particles & yes & no & real \\ \hline
magnetic monopoles & no & yes & hypothetical \\ \hline
dyons & yes & yes & hypothetical \\ \hline
\end{tabular}%
\end{center}
\caption{The four possible particles, observed and hypothetical, in terms of
their electromagnetic charges.}
\label{tab0}
\end{table}
We can now see that this particle classification view has given rise to the
well motivated hypothetical objects, namely magnetic monocles charged only
magnetically, and dyons carrying both the electric and the magnetic charges
associated with the electromagnetic duality in higher dimensions. These
objects are predicted by most grand unified theories in high energy physics
and are now one of the active field of research in many theoretical and
experimental works \cite{13}. As we shall see in what follows, such a view
can be considered as a part of a larger concept being of great utility to
classify physical object families.

\section{Stringy dyonic solutions}

In lower dimensional string compactifications, a certain number of dyonic
objects can be generated. However, these solutions could be obtained from
the geometry of a compact real manifold $X^{n}$ on which higher dimensional
theories are compactified. The geometric information of such
compactifications can be deployed to construct such dyonic objects as
doublets carrying both electric and magnetic charges \cite{15}. Indeed, they
can be generally arranged as
\begin{equation}
\begin{pmatrix}
p \\[3mm]
q%
\end{pmatrix}%
.
\end{equation}%
It is worth noting that now $p$ and $q$ denote the electric and the magnetic
solitonic objets, with $p$ and $q$ dimensions respectively, forming the
dynoic solutions. To see how this could be constructed, consider a gauge
field $C_{p+1}$ coupled to a $p$-brane associated with the field strengtht $%
F_{p+2}=dC_{p+1}$. The corresponding electric charge $Q_{e}$ can be computed
using the integration over a $S^{p+2}$ ($(p+2) $-cycle)
\begin{equation}
Q_{e}=\int_{(p+2)-cycle}F_{p+2}.
\end{equation}%
However, the magnetic charge, associated with the magnetic object, is
calculated via the duality of the field strengtht $F_{p+2}$ in $d$%
-dimensional space time over a $S^{d-p-2}$ ($(d-p-2)$-cycle)
\begin{equation*}
P_{m}=\int_{(d-p-2)-cycle}\ast dC_{p+1}.
\end{equation*}%
To build a dyonic object, one put together the electric and  the magnetic objects
associated with the electromagnetic duality
\begin{equation}
F\longleftrightarrow \ast F.
\end{equation}%
In string theory, when a magnetic dual of an electric $p$-brane is a $q$-brane, where $q$ is related to $p$ by the constraint relation
\begin{equation}
p+q=6-n.
\end{equation}%
This includes two different kinds of the dyonic objects. We refer to them as
fundamental dyonic solutions ($p=q$) and non-fundamental dyonic solutions ($%
p\neq q$), respectively. The first one gives rise to the ordinary dyonic
solution required by
\begin{equation}
p=q=3-\frac{n}{2},
\end{equation}%
which allows to present any fundamental dyonic state in the form of a
doublet of electromagnetic charges with same dimension
\begin{equation}
\begin{pmatrix}
p \\
p%
\end{pmatrix}%
_{p=3-\frac{n}{2}}.
\end{equation}%
This case is analogue of what we call pure state involving only the charged
objects of the same spatial dimension appearing in even dimensional
space-time being  a compact one. It is recalled that one has
three families of dyonic objects such as a dyonic particle in a four
dimensional spacetime, a dyonic string in a six dimensional space-time and a
dyonic membrane in an eight dimensional space-time given respectively by
\begin{equation}
\begin{pmatrix}
0 \\
0%
\end{pmatrix}%
,\quad
\begin{pmatrix}
1 \\
1%
\end{pmatrix}%
,\quad
\begin{pmatrix}
2 \\
2%
\end{pmatrix}%
\end{equation}%
The second family, defined by $p\neq q$, are presented by the following
constraints
\begin{equation*}
q=6-n-p,\qquad p\neq 3-\frac{n}{2}.
\end{equation*}%
In this way, these dyonic objects can be formed by two different branes in
any compact space. Such dyonic objects are presented by a doublet of the
couple ($p,q$)
\begin{equation}
\begin{pmatrix}
p \\
q%
\end{pmatrix}%
_{p+q=6-n}.
\end{equation}%
The unusual solutions can be built in terms of the ten dimensional
space-time theory either from type II superstrings or heterotic brane
doublets
\begin{equation}
\begin{pmatrix}
D0 \\
D6%
\end{pmatrix}%
,\quad
\begin{pmatrix}
D2 \\
D4%
\end{pmatrix}%
,\quad
\begin{pmatrix}
D1 \\
D5%
\end{pmatrix}%
,\quad
\begin{pmatrix}
D3 \\
D3%
\end{pmatrix}%
,\quad
\begin{pmatrix}
NS1 \\
NS5%
\end{pmatrix}%
.
\end{equation}%
These configurations of dyonic states represent the maximally number of
dyonic objects in ten dimensions \cite{15}. It turns out that the number of
such dyonic objects will be decreased by lowering the dimensional space-time
obtained from the compactification $X^{n}$.

Now, we are in position to discuss explicit models from concrete
compactifications. Working with M-theory/superstring inspired models in $d$
dimensions, we consider the compactification on a $n$-dimensional compact
manifold product of the identical manifolds of dimensions $m$ such that
\begin{equation}
n=km.
\end{equation}%
We refer to such manifolds as
\begin{equation}
X^{n}=\underbrace{ Y^{m}\times \ldots \times Y^{m}}_{k-time}
\end{equation}%
where $Y^{m}$ is a $m$-dimensional manifold compact space. In connection
with such theories in the presence of brane solitonic objects, the $d$-dimensional space-time geometry can be split as follows
\begin{equation}
AdS_{p+2}\times \mathbb{S}^{d-p-2-n}\times X^{n}
\end{equation}%
To make contact with dyonic solutions certain conditions should be imposed
on $Y^{m}$. A close inspection shows that $Y^{m}$ should satisfy the
following conditions
\begin{equation}
\left\{ \begin{aligned} b_{0}(Y^m)=1 &\\ b_{i}(Y^m)=0 &&&& 1<i<m-1\\
b_{m}(Y^m)=1 \end{aligned}\right.
\end{equation}%
where $b_{m}(r)$ denote the associated Betti numbers. Performing such
compactifications, we can build dyonic solutions from the brane doublets in $%
d$-dimensional inspired stringy models. Forgetting for while the connection
with string theory models, the compactification can produce $2^{k-1}$ dyonic
double states carrying both electric and magnetic charges. These states are
obtained from branes wrapping $a$-cycles $C_{a}$ in $X$-manifolds. These
cycles are given in terms of the volume forms $\omega_i$ of $Y^{m}$%
-manifolds. Indeed, dyonic stringy solutions, involving the electric and the
magnetic charges, allow one to consider two indices associated with binary
numbers. Using such a property, the cycles can be denoted by
\begin{equation}
C_{a}\equiv C_{e_{1}\ldots e_{k}}.
\end{equation}%
where $e_{i}$ is a binary number taking either 0 or 1. The associated volume
form are given by
\begin{equation}
(\omega _{1})^{e_{1}}\wedge \ldots \wedge (\omega _{k})^{e_{k}}
\end{equation}%
such that
\begin{equation}
\int_{C_{e_{1}\ldots e_{k}}}(\omega _{1})^{e_{1}^{\prime }}\wedge \ldots
\wedge (\omega _{k})^{e_{k}^{\prime }}=\delta _{e_{1}}^{e_{1}^{\prime
}}\ldots \delta _{e_{k}}^{e_{k}^{\prime }}.
\end{equation}%
The dyonic doublet states can be obtained from D-branes wrapping cycles dual
to the following doublet volume form
\begin{equation}
\begin{pmatrix}
p \\
q%
\end{pmatrix}%
\leftrightarrow
\begin{pmatrix}
\omega _{1}^{e_{1}}\wedge \ldots \wedge \omega _{k}^{e_{k}} \\
\omega _{1}^{\overline{e_{1}}}\wedge \ldots \wedge \omega _{k}^{\overline{%
e_{k}}}%
\end{pmatrix}.
\end{equation}%
These dyonic states correspond to $D$-branes with charges
\begin{equation}
\left( Q^{e_{1}\ldots e_{k}},P^{\overline{e_{1}}\ldots \overline{e_{k}}%
}\right)
\end{equation}%
which can be obtained by the following integration
\begin{equation}
Q^{e_{1}\ldots e_{k}}=\int_{C_{e_{1}\ldots e_{k}}}F^{2},\qquad P^{\overline{%
e_{1}}\ldots \overline{e_{k}}}=\int_{C_{\overline{e_{1}}\ldots \overline{%
e_{k}}}}\ast F^{2}.
\end{equation}%
To provide a concrete model, we consider the following compact manifold ${X}%
^{2k}$ where $n=2k$
\begin{equation}
X^{2k}=\underbrace{\mathbf{\mathbb{S}}^{2}\times \mathbf{\mathbb{S}}%
^{2}\times \ldots \times \mathbf{\mathbb{S}}^{2}}_{k}
\end{equation}%
where $\mathbf{\mathbb{S}}^{2}$ is 2-dimensional real sphere. Since $\mathbf{%
\mathbb{S}}^{2}$ is isomorph to $\mathbb{CP}^{1}$, it is useful to consider
the complex geometry. In this way, we take the following factorisation
\begin{equation}
X^{2k}=\underbrace{\mathbf{\mathbb{CP}}^{1}\times \mathbf{\mathbb{CP}}%
^{1}\times \ldots \times \mathbf{\mathbb{CP}}^{1}}_{k}
\end{equation}%
According \cite{110,111}, a nice way to understand such a geometry is to
exploit the $N=2$ sigma model by embedding the involved compact manifold in
local Calabi-Yau manifolds given by
\begin{equation}
\mathcal{O}(-2,\ldots ,-2)\rightarrow \underbrace{\mathbf{\mathbb{CP}}%
^{1}\times \mathbf{\mathbb{CP}}^{1}\times \ldots \times \mathbf{\mathbb{CP}}}%
_{k}.
\end{equation}%
To get the associated cycles, one can use the result of the trivial
fibration of two spaces $M\times N$. According to \cite{130}, the needed
Hodge numbers can be obtained by the following identity
\begin{equation}
h^{(p,q)}(M\times N)=\sum_{\substack{ u+r=p  \\ v+s=q}}%
h^{(u,v)}(M)h^{(r,s)}(N).  \label{ident}
\end{equation}%
It is recalled that the Hodge diagrams associated with $k=1,2,3$ are listed
in Tab.\ref{tab1}.
\begin{table}[th]
\begin{tabular}{|l|l|l|}
\hline
$k=1$ & $%
\begin{tabular}{lllllll}
&  &  & $h^{0,0}$ &  &  &  \\
&  & $h^{1,0}$ &  & $h^{0,1}$ &  &  \\
&  &  & $h^{1,1}$ &  &  &
\end{tabular}%
$ & $%
\begin{tabular}{lllllll}
&  &  & $1$ &  &  &  \\
&  & $0$ &  & $0$ &  &  \\
&  &  & $1$ &  &  &
\end{tabular}%
$ \\ \hline
$k=2$ & $%
\begin{tabular}{lllllll}
&  &  & $h^{0,0}$ &  &  &  \\
&  & $h^{1,0}$ &  & $h^{0,1}$ &  &  \\
& $h^{2,0}$ &  & $h^{1,1}$ &  & $h^{0,2}$ &  \\
&  & $h^{2,1}$ &  & $h^{1,2}$ &  &  \\
&  &  & $h^{2,2}$ &  &  &
\end{tabular}%
$ & $%
\begin{tabular}{lllllll}
&  &  & $1$ &  &  &  \\
&  & $0$ &  & $0$ &  &  \\
& $0$ &  & $2$ &  & $0$ &  \\
&  & $0$ &  & $0$ &  &  \\
&  &  & $1$ &  &  &
\end{tabular}%
$ \\ \hline
$k=3$ & $%
\begin{tabular}{lllllll}
&  &  & $h^{0,0}$ &  &  &  \\
&  & $h^{1,0}$ &  & $h^{0,1}$ &  &  \\
& $h^{2,0}$ &  & $h^{1,1}$ &  & $h^{0,2}$ &  \\
$h^{3,0}$ &  & $h^{2,1}$ &  & $h^{1,2}$ &  & $h^{0,3}$ \\
& $h^{3,1}$ &  & $h^{2,2}$ &  & $h^{1,3}$ &  \\
&  & $h^{3,2}$ &  & $h^{2,3}$ &  &  \\
&  &  & $h^{3,3}$ &  &  &
\end{tabular}%
$ & $%
\begin{tabular}{lllllll}
&  &  & $1$ &  &  &  \\
&  & $0$ &  & $0$ &  &  \\
& $0$ &  & $3$ &  & $0$ &  \\
$0$ &  & $0$ &  & $0$ &  & $0$ \\
& $0$ &  & $3$ &  & $0$ &  \\
&  & $0$ &  & $0$ &  &  \\
&  &  & $1$ &  &  &
\end{tabular}%
$ \\ \hline
\end{tabular}%
\caption{Hodge diagrams}
\label{tab1}
\end{table}
It has been revealed that the invariant volume forms belong to the
cohomology class $H_{+}^{j,j}$ formed by $\prod_{\ell =1}^{j}dz_{\ell
}\wedge d\overline{z_{\ell }}$ where $
w_{\ell }=dz_{\ell }\wedge d\overline{z_{\ell }}$
denotes the volume form associated with $\ell $-th space. It is easy to
calculate that the corresponding Hodge numbers are
\begin{equation}
\dim~H_{+}^{j,j}=h_{+}^{j,j}=\frac{k!}{j!(k-j)!}.
\end{equation}%
It is clear that one has the following relation
\begin{equation}
\dim~H(X^{n})=\sum_{j=0}^{k}h_{+}^{j,j}=2^{k},
\end{equation}%
associated with the total cohomology class.

\section{Tensor network representation of dyonic solutions}

In this section, we would like to investigate dyonic objects using tensor
network formalism \cite{BR,DR}. In particular, we exploit such a formalism
to unveil certain dyonic object properties. It is recalled that usually
tensors are complex having order $k$ and size $L$ \cite{tensor1,tensor2}.
For any tensor, it is interesting to control  the associated degrees of
freedom as well as its organization. Such data could be used to encode
certain physical quantities corresponding to fundamental and non fundamental
objects including dyonic ones. For later use, we will consider the  real tensors
with a fixed size being $L=2$.

Roughly speaking, the tensors, which  will be used here, are defined as a series of
real numbers labeled by $k$ indices associated with legs in graph
representattions. In such a context, a scalar, which is one number and
labeled by zero index, is a $0$th-order tensor. Graphically, we represent a
scalar by dot as illustrated in Fig.1.
\begin{figure}[tbph]
\centering
\includegraphics[scale=0.3]{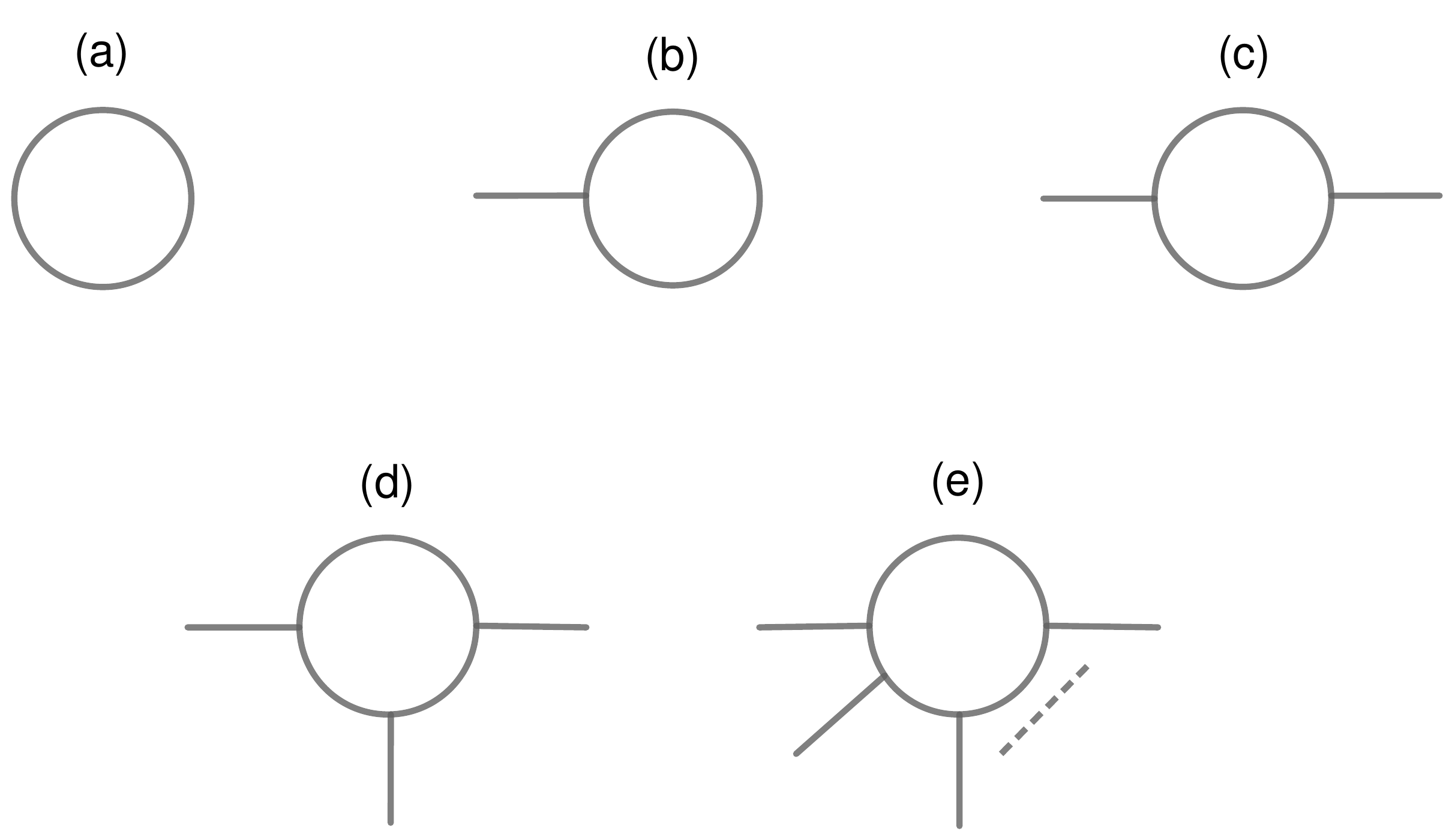}
\caption{Tensor representation. (a) scalar, (b) vector, and (c-d-e) tensors
associated with polyvalent vertices}
\end{figure}

A $2$-component vector consists of $2$ real numbers labeled by one index,
being a $1$st-order tensor. In this way, one can write the state vector of a
spin-$1/2$ as follows
\begin{equation}
|\Psi \rangle =C_{1}|{0}\rangle +C_{2}|{1}\rangle =\sum_{e=0,1}C_{e}|{e}%
\rangle ,
\end{equation}%
where the coefficients $C_{i}$ correspond to  a two-component vector.
It is recalled that $|{0}\rangle $ and $|{1}\rangle $ are usually used to
represent spin up and down states. Graphically, we use a dot with one open
bond to represent such a vector (see (a) in figure 1). However, a matrix is
in fact a $2$nd-order tensor. Considering two spins as an example, the state
vector can be written under an irreducible representation as a
four-dimensional vector. Using the local basis of each spin associated with
binary notation, the state can be written as follows
\begin{equation}
|\Psi \rangle =C_{00}|{0}{0}\rangle +C_{01}|{0}{1}\rangle +C_{10}|{1}{0}%
\rangle +C_{11}|{1}{1}\rangle
=\sum_{e_{1}e_{2}=0}^{1}C_{e_{1}e_{2}}|e_{1}e_{2}\rangle ,
\end{equation}%
where $C_{e_{1}e_{2}}$ represent now a matrix with two indices. It is worth
noting that the difference between a $(2\times 2)$ matrix and a $2^{2}$%
-component vector is just the way of labeling the tensor elements.
Graphically, we use a dot with two bonds to represent a matrix and its two
indices (Figure 1).\newline
It is then natural to define an $k$-th order tensor. Considering $k$ spins,
the $2^{k}$ coefficients can be written as a $k$-th order tensor $C$, which
satisfies
\begin{equation}
|\Psi \rangle =\sum_{e_{1}\ldots e_{k}=0}^{1}C_{e_{1}\ldots e_{k}}|{e_{1}}%
\ldots {e_{k}}\rangle .
\end{equation}%
Graphically, such a $k$-th order tensor is represented by a dot connected
with $k$ open bonds (polyvalent vertex of order $k$, Figure 1). Now we can
define Tensor Network (TN), as the contraction of many tensors. A TN is a
set of tensors where some, or all, of its indices are contracted according
to some pattern. Contracting the indices of a TN is called, for simplicity,
contracting the TN. In general, the contraction of a TN with some open
indices gives as a result another tensor, and in the case of not having any
open indices the result is a scalar.\newline

Having given a concise review on tensors, we move to make contact with
dyons. Indeed, a close inspection, in the above concrete stringy  models,
shows that the dyonic state can be associated with a tensor $T_{e_{1}\ldots
e_{k}}$ defined by the state
\begin{equation}
|\Psi \rangle =\sum_{e_{i}=0,1}T_{e_{1}\ldots e_{k}}|{e_{1}\ldots e_{k}}%
\rangle ,
\end{equation}%
where the vector $|{e_{1}\ldots e_{k}}\rangle $ is given by $|{e_{1}\ldots
e_{k}}\rangle =|{e_{1}}\rangle \otimes \ldots \otimes |{e_{k}}\rangle $.%
\newline
To give a tensor network realisation of such dyonic objects, we introduce a
new tensor defined by
\begin{equation}
T_{e_{1}\ldots e_{k}}\rightarrow \bar{T}_{e_{1}\ldots e_{k}}=T_{\bar{e}%
_{1}\ldots \bar{e}_{k}}  \label{dioniccond}
\end{equation}%
such that $e_{i}+\bar{e}_{i}=1$.\newline
In this way, the above state can be written as
\begin{equation}
|{\psi }\rangle =\sum_{e_{i}}T_{e_{1}\ldots e_{k}}|{e_{1}\ldots e_{k}}%
\rangle +\sum_{\bar{e}_{i}}T_{\bar{e}_{1}\ldots \bar{e}_{k}}|{\bar{e}%
_{1}\ldots \bar{e}_{k}}\rangle
\end{equation}%
where the tensor values correspond to the involved charges. Indeed, one
proposes
\begin{equation}
\begin{split}
T_{e_{1}\ldots e_{k}}& =Q_{e_{1}\ldots e_{k}} \\
T_{\bar{e}_{1}...\bar{e}_{k}}& =P_{\bar{e}_{1}\ldots \bar{e}_{k}}
\end{split}%
\end{equation}%
subject to $e_{i}+\bar{e}_{i}=1$. In this way, the degrees of freedom of $T$
can be split as
\begin{equation}
2^{k}=2^{k-1}+2^{k-1}.
\end{equation}%
Now, we are in position to build a new tensor carrying information on dyonic
objects by combing $T$ and $\bar{T}$ tensor defined previously. The
suggested notation is given by
\begin{equation}
X_{\bar{e}_{1}\ldots \bar{e}_{k}}^{e_{1}\ldots e_{k}}=%
\begin{pmatrix}
T^{e_{1}\ldots e_{k}} \\
T_{\bar{e}_{1}\ldots \bar{e}_{k}}%
\end{pmatrix}
\label{dionfield}
\end{equation}%
In this tensor realization, the dyonic state can be expressed as as
\begin{align}
|{\psi }\rangle& =\sum_{e_{i}+\bar{e}_{i}=1}X_{\bar{e}_{1}\ldots \bar{e}%
_{k}}^{e_{1}\ldots e_{k}}|{e_{1}\bar{e}_{1}}\rangle \otimes \ldots \otimes |{%
e_{k}\bar{e}_{k}}\rangle \\
& =\sum_{e_{i}+\bar{e}_{i}=1}X_{\bar{e}_{1} \ldots \bar{e}_{k}}^{e_{1}\ldots
e_{k}}\prod_{i}\otimes _{i}|{e_{i}\bar{e}_{i}}\rangle .
\end{align}%
Graphically, this state can be illustrated by the tensor given in Fig.2
\begin{figure}[tbph]
\centering
\includegraphics[scale=0.3]{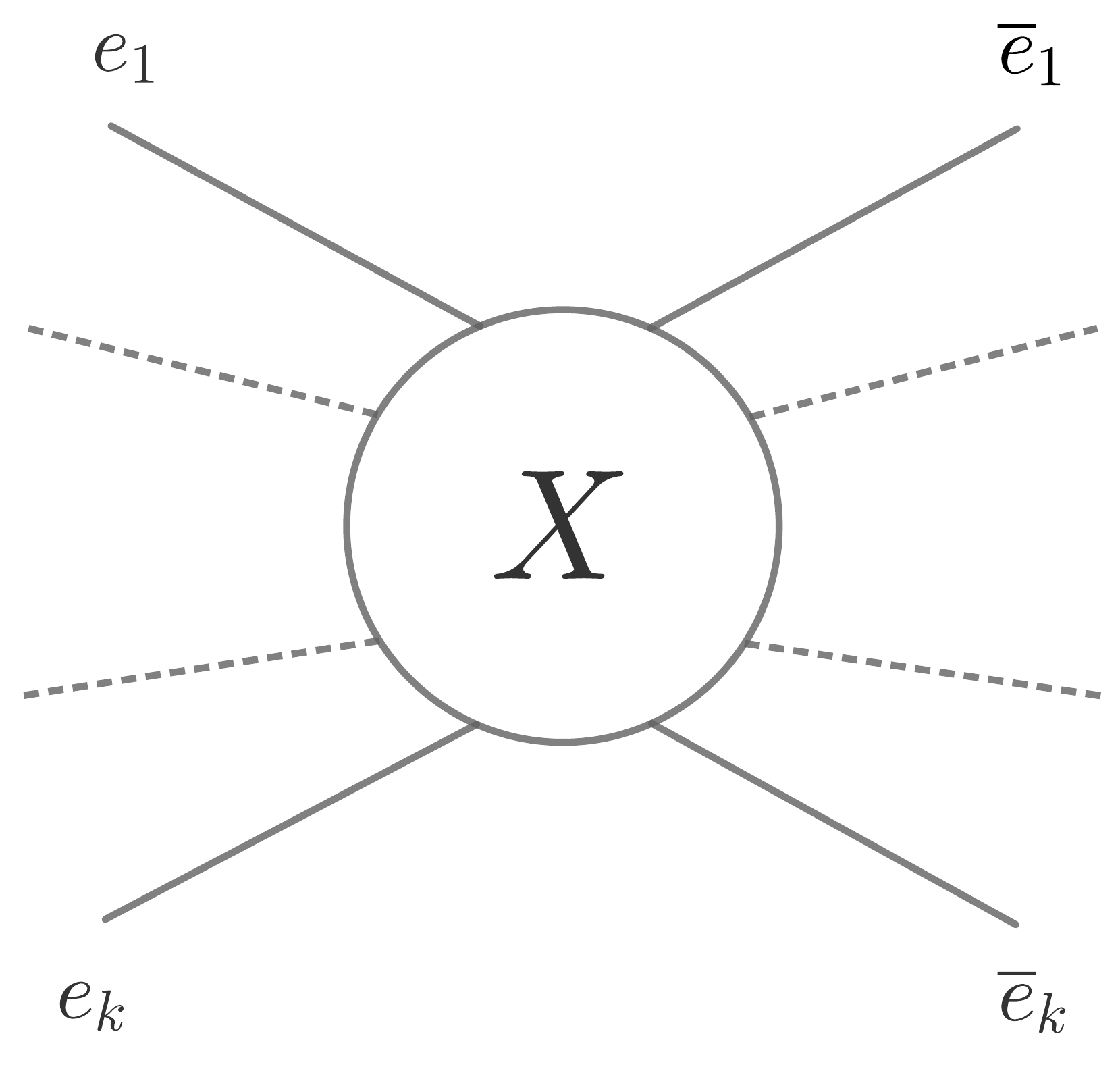}
\caption{Tensor state representation}
\end{figure}

Roughly, we expect that the theory that supports these objects is a color
tensor model with a gauge group $U(2)^{\otimes k}$ which acts on each slot
by complex rotation. In fact, multiparticle states are given by linear
combinations of the full contraction of $r$ copies of the tensor $T$ with $r$
copies of the tensor $\overline{T}$, providing that the states are gauge
invariant. To simplify notation, consider a tensor with three indices,
namely $k=3$. The results can be generalized straightforwardly. For one
particle, we have
\begin{equation}
\mathcal{O}^{(1)}=T_{ijk}\overline{T}^{ijk},
\end{equation}
which is the unique scalar that can be constructed (remember that in color
tensor models the contractions must always happen on indices of the same
slot). For two indices, there are four choices given by
\begin{eqnarray}
\mathcal{O}_1^{(2)}&=&T_{i_1j_1k_1}T_{i_2j_2k_2}\overline{T}^{i_1j_1k_1}%
\overline{T}^{i_2j_2k_2},\quad \mathcal{O}_2^{(2)}=T_{i_1j_1k_1}T_{i_2j_2k_2}%
\overline{T}^{i_2j_1k_1}\overline{T}^{i_1j_2k_2}  \notag \\
\mathcal{O}_3^{(2)}&=&T_{i_1j_1k_1}T_{i_2j_2k_2}\overline{T}^{i_1j_2k_1}%
\overline{T}^{i_2j_1k_2},\quad \mathcal{O}_4^{(2)}=T_{i_1j_1k_1}T_{i_2j_2k_2}%
\overline{T}^{i_1j_1k_2}\overline{T}^{i_2j_2k_1}.
\end{eqnarray}
According to \cite{BR}, $r$ copies we have as many invariants as
\begin{equation}  \label{spanset}
\mathcal{O}_{\alpha\beta\gamma}=T_{i_1j_1k_1}\dots T_{i_rj_rk_r}\overline{T}%
^{i_{\alpha(1)}j_{\beta(1)} k_{\gamma(1)}}\dots\overline{T}%
^{i_{\alpha(r)}j_{\beta(r)} k_{\gamma(r)}}|\quad\alpha,\beta,\gamma \in S_r,
\end{equation}
subject to the equivalence
\begin{equation}
\mathcal{O}_{\alpha\beta\gamma}\sim \mathcal{O}_{\alpha^{\prime
}\beta^{\prime }\gamma^{\prime }}\quad\text{if } \quad \alpha^{\prime
}=\tau\alpha\sigma,\,\, \beta^{\prime }=\tau\beta\sigma,\,\, \gamma^{\prime
}=\tau\gamma\sigma,
\end{equation}
for some $\sigma,\tau\in S_r$. The equivalence takes into account the
freedom to shuffle $T$ and $\overline{T}$ slots in \eqref{spanset}. As a
comment, note that the same invariants could have been constructed with $r$
copies of \eqref{dionfield} and patterns of contraction. We are expressing
the same, although we believe that this construction is simpler.\newline

Besides, for $L< r$ (which is going to be the case since $L=2$), the number
of invariants gets reduce in account of redundancies due to the small number
of degrees of freedom. Following \cite{DR,Re1,Re2,Re3}, the exact number of invariants
has been computed and is
\begin{equation}  \label{inv}
\text{$\sharp$ of invariants}=\sum_{\substack{ \mu,\nu,\lambda \vdash r  \\ %
l(\mu),l(\nu),l(\lambda)\leq 2}}g_{\mu\nu\lambda}^2,
\end{equation}
where $g_{\mu\nu\lambda}$ are the Kronecker coefficients labeled by three
Young diagrams or partitions of $r$, and the fact that $L=2$ translates into
the number of the rows of each Young diagram being equal or less than 2, as
indicated in the sum \eqref{inv}.

It is remarked that Kronecker coefficients are hard to compute in general.
However the restriction on the partitions to have at most two rows should
make a dramatic simplification. This case could be studied somewhere and
some neat formulas found for the counting. It would be interesting to find
it.

Now, we could interpret \eqref{dioniccond} as the dyonic condition. It is
actually a restriction in the number of degrees of freedom since given $%
T_{ijk}$ one immediately knows what is $T_{\bar{\imath}\bar{j}\bar{k}}$,
which is the conjugate number. It is noted that, thanks to \eqref{dioniccond}%
, the invariants are real since
\begin{eqnarray}
\overline{\mathcal{O}}_{\alpha \beta \gamma } &=&\overline{%
T_{i_{1}j_{1}k_{1}}\dots T_{i_{r}j_{r}k_{r}}\overline{T}^{i_{\alpha
(1)}j_{\beta (1)}k_{\gamma (1)}}\dots \overline{T}^{i_{\alpha (r)}j_{\beta
(r)}k_{\gamma (r)}}}  \notag \\
&=&\overline{T}_{i_{1}j_{1}k_{1}}\dots \overline{T}_{i_{r}j_{r}k_{r}}T^{i_{%
\alpha (1)}j_{\beta (1)}k_{\gamma (1)}}\dots T^{i_{\alpha (r)}j_{\beta
(n)}k_{\gamma (r)}}  \notag \\
&=&T_{\bar{\imath}_{1}\bar{j}_{1}\bar{k}_{1}}\dots T_{\bar{\imath}_{r}\bar{j}%
_{r}\bar{k}_{r}}\overline{T}^{\bar{\imath}_{\alpha (1)}\bar{j}_{\beta (1)}%
\bar{k}_{\gamma (1)}}\dots \overline{T}^{\bar{\imath}_{\alpha (r)}\bar{j}%
_{\beta (r)}\bar{k}_{\gamma (n)}}  \notag \\
&=&T_{i_{1}j_{1}k_{1}}\dots T_{i_{r}j_{r}k_{r}}\overline{T}^{i_{\alpha
(1)}j_{\beta (1)}k_{\gamma (1)}}\dots \overline{T}^{i_{\alpha (r)}j_{\beta
(r)}k_{\gamma (r)}}=\mathcal{O}_{\alpha \beta \gamma }.
\end{eqnarray}

 After improving the  comprehension of the mathematical structure of the
matrix product states, for instance the TN, one might wonder why not use
such a method  to advance in the mathematical foundations of related physics. This is
possible thanks to the properties, especially the encoded symmetries, of the
TN states in many-body systems. Indeed,  the TNs could be employed to explore  results
in condensed matter physics and quantum information theory. More precisely, this includes
the microscopic origin of magnetism with spin systems and quantum computing simulations.
For many-body systems, the spectral resolution of the corresponding
acting Hamiltonian is always a highly complex task due to  the associated
large Hilbert space. The latter  grows, exponentially,  with the number of system  particles. Such a fast growth is a relevant characteristic for  certain
physical behaviors such as phase transitions and quantum computation investigations.

Although  it seems that the TNs have not been mastered and exploited well
enough, we believe that they are a powerful mathematical framework to deal
with certain deepest physical phenomena.

\section{Conclusion}

In this paper, we have investigated stringy dyonic objects and tensor
network formalism. Inspired and motivated by known results obtained from
non-trivial theories including string theory, we have provided shortly the
origin and certain related concepts. After that, we have elaborated a
general framework of dyonic solutions in the brane structure and show how
are naturally obtained in various dimensions within stringy
compactifications on a concrete complex geometry. Then, we have discussed
dyonic solution representations using tensor network formalism by revealing
some characteristics associated with extra dimensions.

Assuming the existence of such dimensions, we believe that such a
theoretical investigation could shed, somehow, some light on experimental
searches for dyons, through either their direct production or indirect
detection benchmarks at the current and future accelerators.

This work comes with certain open questions. A naturel question is make
contact with MERA being a kind of tensor network states explored in the
study of black holes  \cite{blackhole}.
We try to address such question elsewhere.

\section*{Acknowledgments}
The authors,  being listed in alphabetical  order,  would like to thank P. Diaz, A. Marrani, E. Torrente-Lujan,  A. Segui and Y. Sekhmani for discussions on related topics.  They are also grateful to the anonymous referee for their careful reading of our manuscript, insightful comments, and suggestions, which have allowed us to improve this paper. This work is partially supported by the ICTP through AF-13.

\end{document}